\documentclass[
aps,prd,
showpacs,twocolumn,notitlepage,
amssymb,amsmath,amsfonts,
nofootinbib,superscriptaddress,
floats,
amsmath,amssymb,
aps,
]{revtex4-2}

\usepackage{graphicx}
\usepackage{xcolor}
\usepackage[colorlinks=true]{hyperref}
\hypersetup{citecolor=cyan,linkcolor=magenta}
\usepackage{multirow,array}

\usepackage{amsmath,amssymb,siunitx}
\usepackage{hyperref}
\usepackage{rotating}
\usepackage{ulem}
\usepackage{array}
\usepackage{color}
\usepackage{multirow}
\usepackage{float}

\usepackage{comment}

\begin{document}
\title{Jet from binary neutron star merger with prompt black hole formation}

\author{Kota Hayashi}
\affiliation{Max Planck Institute for Gravitational Physics (Albert Einstein Institute),
Am M{\"u}hlenberg 1, Postdam-Golm 14476, Germany}
\affiliation{Center for Gravitational Physics and Quantum Information, Yukawa Institute for Theoretical Physics,
  Kyoto University, Kyoto 606-8502, Japan}

\author{Kenta Kiuchi}
\affiliation{Max Planck Institute for Gravitational Physics (Albert Einstein Institute),
Am M{\"u}hlenberg 1, Postdam-Golm 14476, Germany}
\affiliation{Center for Gravitational Physics and Quantum Information, Yukawa Institute for Theoretical Physics,
  Kyoto University, Kyoto 606-8502, Japan}

\author{Koutarou Kyutoku}
\affiliation{Department of Physics, Graduate School of Science, Chiba University, Chiba 263-8522, Japan}
\affiliation{Interdisciplinary Theoretical and Mathematical Sciences Program (iTHEMS), RIKEN, Wako, Saitama 351-0198, Japan 
}

\author{Yuichiro Sekiguchi}
\affiliation{Department of Physics, Toho University, Funabashi, Chiba 274-8510, Japan}

\author{Masaru Shibata}
\affiliation{Max Planck Institute for Gravitational Physics (Albert Einstein Institute),
Am M{\"u}hlenberg 1, Postdam-Golm 14476, Germany}
\affiliation{Center for Gravitational Physics and Quantum Information, Yukawa Institute for Theoretical Physics,
  Kyoto University, Kyoto 606-8502, Japan}
  
\date{\today}

\begin{abstract}
We performed the longest numerical-relativity neutrino-radiation magnetohydrodynamics simulation for a binary neutron star merger that extends to $\approx1.5\mathrm{\,s}$ after the merger.
We consider the binary model that undergoes the prompt collapse to a black hole after the merger with asymmetric mass 1.25$\,M_{\odot}$ and 1.65$\,M_{\odot}$ and SFHo equation of state.
We find the Poynting flux-driven collimated outflow as well as the gravitational wave emission, neutrino emission, dynamical mass ejection, and post-merger mass ejection facilitated by magnetorotational instability-driven turbulent viscosity in a single self-consistent binary neutron star merger simulation. 
A magnetosphere dominated by the aligned global magnetic field penetrating the black hole develops along the black-hole spin axis after the turbulence in the remnant disk is enhanced. A jet with the Poynting flux with isotropic-equivalent luminosity of $\sim10^{49}\mathrm{\,erg/s}$ is launched, and the duration of the high luminosity is expected to be $O(1)$~s.
\end{abstract}

\maketitle

{\it Introduction}.--
Multi-messenger astrophysics, including gravitational wave (GW) detection, began with the observation of GW170817, AT2017gfo, and GRB170817A from a binary neutron star (BNS) merger~\cite{abbott2017oct1,abbott2017oct2,abbott2017oct3,Goldstein2017oct,Savchenko2017oct,Mooley2018sep}.  
The GWs constrain the nuclear equation of the state (EOS) of neutron star (NS) matter~\cite{abbott2017oct1,Abbott2018oct,Abbott2019jan,De:2018uhw,Narikawa:2019xng}.
The kilonova emission indicated that neutron-rich matter is likely ejected from the system~\cite{lattimer1974,eichler1989,li1998nov,metzger2010jun}, with the synthesis of heavy elements by the $r$-process (rapid neutron capture process)~\cite{Tanaka:2013ana,Wanajo2014jul,J-GEM:2017tyx,Tanaka:2017qxj,Domoto:2021xfq,Domoto:2022cqp}.
The detection of a short-hard gamma-ray burst (SGRB) indicated that the relativistic jet was launched from the system~\cite{eichler1989,nakar2007apr,berger2014jun}.

Two years later, GW190425 revealed that the binary system with large total mass $\sim3.4M_{\odot}$ exists~\cite{abbott2020mar}.
The GW signal is consistent with BNS merger (though the heavy component of the system being a black hole (BH) is not ruled out~\cite{kyutoku2020feb}).
The total mass of the system is larger than the BNS systems known~\cite{lorimer2008nov,Farrow:2019xnc} and it is expected to experience a prompt collapse to the BH after the merger~\cite{Shibata2005apr,Shibata2006mar,Kiuchi2009sep,Kiuchi2010apr,bauswein2013sep,Zappa2018mar,Bauswein2020oct,Bauswein2021jun}.
Since this event was a single detector event, the sky localization was poor, and an electromagnetic counterpart associated with this event was not detected~\cite{Dudi:2021abi}.
However, if the mass asymmetry is large, we may expect that the system undergoes the prompt collapse to a BH with a massive disk remaining~\cite{shibata2003oct,Shibata2006mar}.
If the GW event from such a source is detected with sufficient localization, the associated electromagnetic counterparts could be expected, as was the case for GW170817.

The theoretical modeling of the entire evolution for the prompt-collapse BNS merger and the prediction of the multi-messenger signals are highly demanded to interpret the foreseen observation.
For this purpose, it is crucial to perform numerical relativity simulations incorporating magnetohydrodynamics and neutrino-radiation transfer.
Further, the simulation has to cover an entire evolution stage extending  $O(1)\mathrm{\,s}$ for the post-merger stage in order to self-consistently explore the mass ejection and jet launch.

Since 2005, BNS merger simulations with magnetohydrodynamics and/or microphysics of the NS matter focusing on the merger and early post-merger stage have been performed~\cite{Aguilera2020nov,Aguilera2022feb,Aguilera2023nov,Anderson2008may,Dionysopoulou2015oct,Giacomazzo2009oct,Giacomazzo2011feb,Kiuchi2014aug,Kiuchi2015dec,Kiuchi2018jun,kiuchi2022dec,Liu2008jul,Most2023apr,Most2023dec,Mosta2020oct,Palenzuela2015aug,Palenzuela2022may,Palenzuela2022jul,Ruiz2017oct,Ruiz2019apr,Ruiz2020mar,Ruiz2021dec,Sun2022may,Sekiguchi2011jul,Sekiguchi2015mar,Sekiguchi2016jun,Radice2016dec,Radice2018dec,Foucart2016dec,Foucart2023may,Zappa2023mar,Lehner2016sep}. 
Also, simulations for BNS merger remnants, focusing on the post-merger stage, have been recently  done~\cite{christie2019sep,Duez2006jan,Duez2006may,fernandez2018oct,kiuchi2012sep,Lehner2012nov,Shibata2006jan,Shibata2011jun,Shibata2021feb,Shibata2021sep,Siegel2013jun,Siegel2018may,Siegel2017dec,Stephens2007jun,fujibayashi2020apr,fujibayashi2020dec,fujibayashi2020sep,Fujibayashi2023jan}.
However, these simulations have limitations in that they were not long enough to explore the entire post-merger evolution, or they employed non-self-consistent initial data.
Only Ref.~\cite{kiuchi2023jul} has successfully simulated the entire evolution of BNS merger self-consistently. 
Concerning the prompt-collapse BNS case, Ref.~\cite{Ruiz2017oct} have performed general relativistic magnetohydrodynamics simulations.
It showed that up to $\sim26\mathrm{\,ms}$ after the merger, there is no evidence for aligned magnetic field formation in the polar region and Poynting flux-driven outflow. 
However, this time scale may not be long enough to explore the jet launch since the event GW170817 and GRB170817A had a time lag of $\approx1.7\mathrm{\,s}$~\cite{abbott2017oct1,abbott2017oct2,abbott2017oct3} and black hole-neutron star (BHNS) merger simulations showed that it requires $200\mathrm{\,ms} \lesssim t \lesssim3\mathrm{\,s}$ after the merger for magnetosphere formation and Poynting flux launching depending on the binary models~\cite{hayashi2022jul,hayashi2023jun}.

In order to delineate the entire picture of the prompt-collapse BNS merger, we performed a second-long numerical-relativity neutrino-radiation magnetohydrodynamics simulation on the Japanese supercomputer Fugaku.
This Letter reports the multi-messenger signals driven by the prompt-collapse BNS merger focusing especially on the launch of the Poynting flux-driven collimated outflow, i.e., jet.

{\it Numerical methods and model}.--
The numerical method used in this work is the same as that summarized in Ref.~\cite{kiuchi2022dec}.
We solve Einstein's equation by a puncture-Baumgarte-Shapiro-Shibata-Nakamura (BSSN) formalism~\cite{shibata1995nov,baumgarte1998dec,campanelli2006mar,baker2006mar,marronetti2008mar} with the Z4c constraint propagation prescription~\cite{hilditch2013oct,Kyutoku:2014yba,kawaguchi2015jul}.
The fourth-order finite differentiation in space and time discretizes the equation.
The magnetohydrodynamics equations are solved using the HLLD Reimman solver~\cite{mignone2009mar} with the constrained transport scheme~\cite{gardiner2008apr} and the divergence-free- and magnetic-flux-preserving mesh refinement scheme~\cite{balsara2009,Neuweiler2024jul_arxiv}. 
Our code implements the gray M1+Leakage scheme for the neutrino radiation transfer to take into account the neutrino cooling and heating~\cite{thorne1981feb,Shibata2011jun,sekiguchi2010aug,sekiguchi2012oct,fujibayashi2020sep}.

The NS matter is modeled by nuclear-theory-based finite-temperature EOS, SFHo~\cite{steiner2013sep}, for a high-density range and Helmholtz EOS~\cite{timmes2000} for a low-density range.
The maximum mass of the isolated non-rotating NS with SFHo is $2.06 \, M_{\odot}$.
The LORENE library~\cite{lorene} is employed to prepare an asymmetric irrotational BNS with mass $M_1=1.25M_{\odot}$ and $M_2=1.65M_{\odot}$ in quasi-equilibrium, assuming the neutrinoless beta-equilibrium cold state, as the initial data.
The initial orbital frequency $\Omega_0$ is set to be $\Omega_0 m_0 G/c^3=0.025$ where $m_0=M_1+M_2$, $c$ the speed of light, and $G$ gravitational constant. 

We initially superimposed a poloidal magnetic field confined inside the NS.
The magnetic field is given in terms of the vector potential as
\begin{eqnarray} \label{init_b_field_p}
  A_j &= & \{ -(y-y_{\mathrm{NS}}) \delta_j^{~x} + (x-x_{\mathrm{NS}}) \delta_j^{~y} \} \nonumber \\
  && \times  A_b \max(P/P_{\mathrm{max}} - 2 \times 10^{-4} , 0)^{0.5},
\end{eqnarray}
where $x_{\mathrm{NS}}$ and $y_{\mathrm{NS}}$ are the coordinate center of each NS, $P$ and $P_{\mathrm{max}}$ are the pressure and its maximum value, respectively, and $A_b$ is constant and set so that the maximum magnetic-field strength becomes $10^{15}$\,G.
This definition is similar to in, e.g., Refs.~\cite{Kiuchi2018jun,kiuchi2023jul}, but differs for the power index 0.5 instead of $2$ used in the previous works. 
We intended both NSs to have a strong magnetic field that extends to their surface. 
With this, a strong magnetic field remains in the massive disk formed after the prompt BH formation. 
The strong initial magnetic field is essential to resolve magnetorotational instability (MRI) in the remnant disk with limited grid resolution.

This simulation is performed on a fixed-mesh refinement domain with 14 refinement levels.
The grid resolution in the finest level is $\Delta x_{14}=150$\,m and the coarser domain has the resolution of $\Delta x_{l}=2 \Delta x_{l+1}$ where $l=1, \cdots ,13$.
The finest domain covers $[-38.1 \, \mathrm{km},38.1 \, \mathrm{km}]^3$ and the whole computational domain is $[-312,115.2 \, \mathrm{km},312,115.2 \, \mathrm{km}]^3$  in the Cartesian coordinates.

We set the atmosphere with the rest-mass density $10^{3}\mathrm{\,g/cm^3}$ for the finest domain and  $10^{3}(r/38.1\mathrm{\,km})^{-3}\mathrm{\,g/cm^3}$ for the coarser domains where $r$ is the radial coordinate.
We assume the constant density once the atmosphere density reaches $\approx0.166\mathrm{\,g/cm^3}$.
We set the temperature and the electron fraction of the atmosphere to be $10^{-3}\mathrm{\,MeV}$ and $0.5$,
respectively. 

The computational cost is about 130 million CPU hours. Each job was carried out using 20,736--82,944 CPUs on Fugaku.

{\it Inspiral and merger}.--
The BNS merger can be characterized by three stages; inspiral, merger, and post-merger stages.
In the present simulation, the system inspirals for five orbits before the merger. In this stage, the magnetic field plays essentially no role since the ratio of the electromagnetic field energy to the internal energy is less than $10^{-6}$.
Since the total mass of the system is high $m_0=2.9\,M_{\odot}$  the merged NSs collapse promptly in the merger stage~\cite{Shibata2005apr,Shibata2006mar,Kiuchi2009sep,bauswein2013sep,bauswein2017nov, Shibata:2017xdx}, forming a BH with its mass $\approx 2.77 {\,M_{\odot}}$ and dimensionless spin $\approx 0.76$.
\footnote{Note that due to the lack of grid points resolving the BH, the BH spin gradually decreases numerically, so we switch to the Cowling approximation at $t-t_\mathrm{merger}\approx0.44\mathrm{\,s}$. 
This can be done safely since the baryonic mass of the disk is less than 1\% of the BH mass at this time.
}

We define the merger time $t_\mathrm{merger}=16.4$\,ms at which the amplitude of GWs becomes maximum.
Since the binary is highly asymmetric, a massive disk is formed around the BH after the merger~\cite{shibata2003oct,Shibata2006mar}. 
The barionic mass of the disk at $t-t_\mathrm{merger}\approx 20\mathrm{\,ms}$ is $\approx6.2 \times 10^{-2} {\,M_{\odot}}$.

\begin{figure}[t]
      \begin{center}
        \includegraphics[scale=0.22]{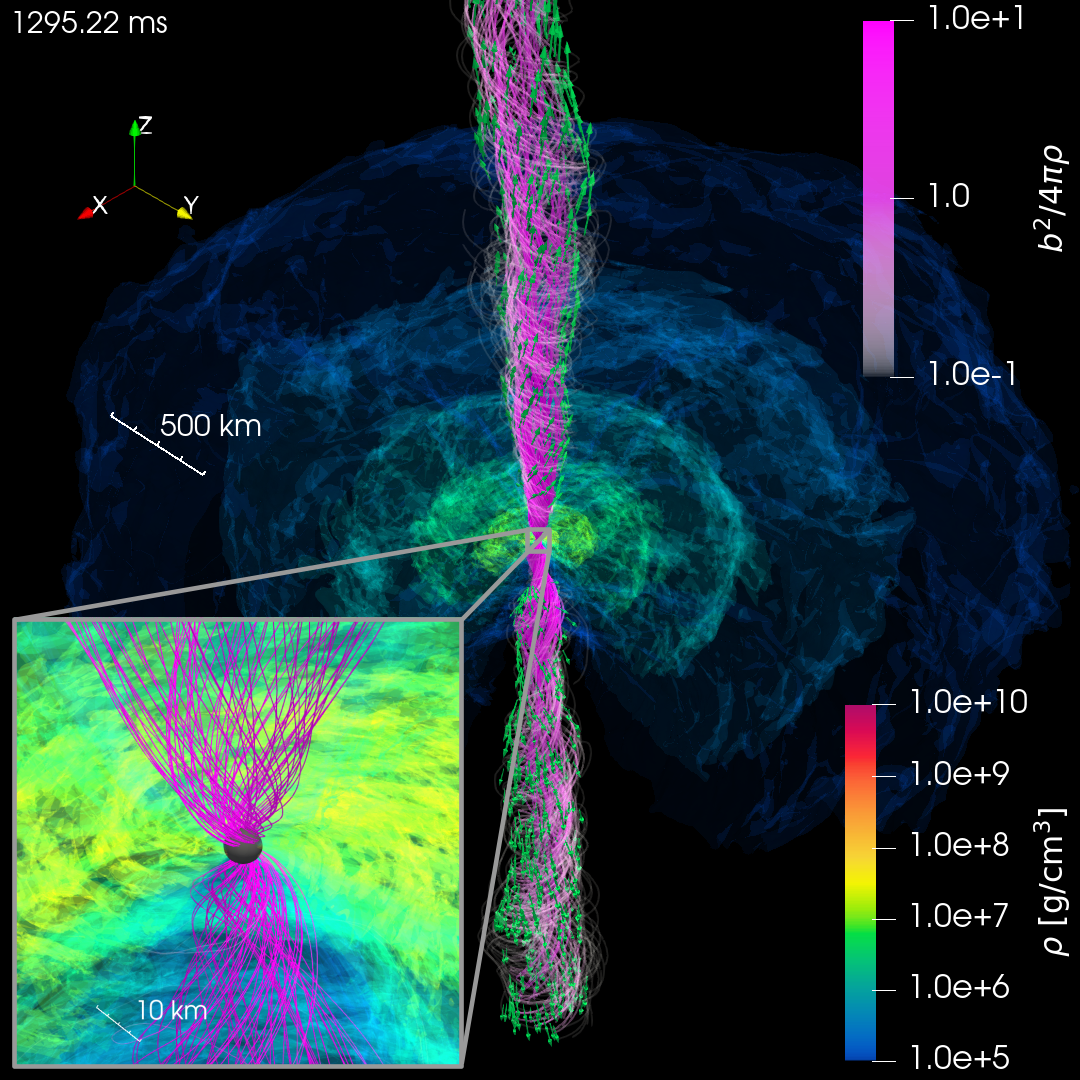} \\
        \caption{The 3D snapshot at $t-t_{\mathrm{merger}}\approx1.3\mathrm{\,s}$ in a domain of $\sim10^3\mathrm{\,km}$ and $\sim10^2\mathrm{\,km}$.
        The rest-mass density (contours), magnetic-field lines penetrating the apparent horizon (lines), and the outflow in the magnetosphere (arrows) are shown.
        The black sphere at the center shows the apparent horizon.
        See also the animation~\cite{3Danim} and interactive 3D snapshot~\cite{interactive3Dsnap}.
        }
        \label{fig:3dsnapshot}
      \end{center}
\end{figure}

{\it Poynting flux-driven outflow}.--
Figure~\Ref{fig:3dsnapshot} displays the 3D snapshot at $t-t_{\mathrm{merger}}\approx1.3\mathrm{\,s}$ in a domain of $\sim10^3\mathrm{\,km}$ and $\sim10^2\mathrm{\,km}$.
The rest-mass density, magnetic-field lines penetrating the apparent horizon, and the outflow in the magnetically dominated region are shown.
We find that along the BH spin axis, the magnetosphere is developed due to the large-scale magnetic field produced by the MRI-driven $\alpha\Omega$ dynamo~\cite{Brandenburg2005oct,ReboulSalze2022nov,kiuchi2024mar}.
In this magnetosphere, we find an outgoing Poynting flux with its isotropic-equivalent luminosity $\approx10^{49}\mathrm{\,erg/s}$ and its duration of $\sim O(1) \mathrm{\,s}$ that could power an SGRB (see below for more details).
The luminosity is lower than the typically observed SGRBs~\cite{nakar2007apr,berger2014jun}, but the duration is consistent.
Below we describe the magnetosphere formation process in detail.

After the merger in the remnant disk, the magnetic field is amplified by the winding~\cite{shibata2006} and MRI~\cite{balbus1991,balbus1998}.
The fastest growing mode of the MRI is sufficiently resolved for $\rho\lesssim10^{10}\,\mathrm{g/cm^3}$ and $t-t_\mathrm{merger}\gtrsim0.02\mathrm{\,s}$, and fully resolved for whole disk for $t-t_\mathrm{merger}\gtrsim0.1\mathrm{\,s}$ (see Supplement Material).
Figure~\ref{fig:mag} plots the azimuthally-averaged toroidal magnetic field along the polar direction at $r \approx 50$\,km.
In addition, the animation~\cite{2Drtanim} displays the azimuthally-averaged rest-mass density, mean poloidal magnetic field, and the MRI quality factor on the $r$--$\theta$ plane.
They show that as the MRI is resolved, a magnetic field with the opposite polarity is periodically generated near the equatorial plane in the disk.
The generated magnetic field ascends to the polar region and accumulates there.
As the magnetic field with the opposite polarity propagates, they cancel out each other and dissipate away due to the reconnection.
The period for the polarity flips at $r\approx50\mathrm{\,km}$ is $\approx0.03$--$0.04\mathrm{\,s}$ which agrees with the period derived by the $\alpha\Omega$ dynamo theory~\cite{Brandenburg2005oct,ReboulSalze2022nov,kiuchi2024mar} $\approx0.03\mathrm{\,s}$.
The detailed analysis of the $\alpha\Omega$ dynamo will be reported in a separate paper~\cite{ReboulSalze_in_prep}.
These facts show that the MRI-driven $\alpha\Omega$ dynamo is activated, and the large-scale magnetic field with a length scale comparable to the size of the disk is generated.

\begin{figure}[t]
      \begin{center}
        \includegraphics[scale=0.5]{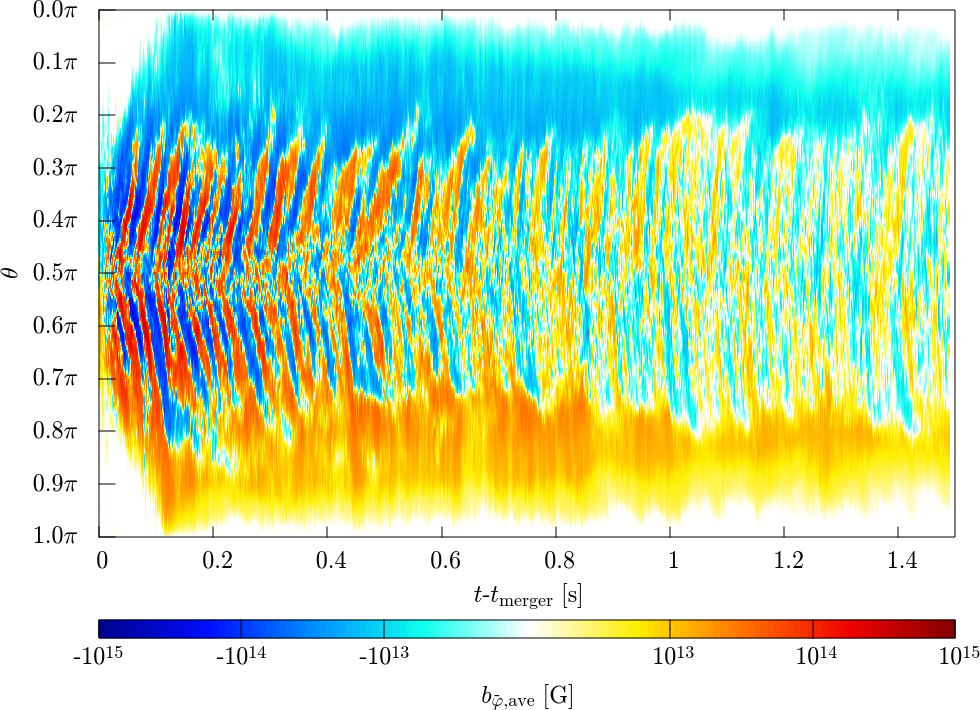} \\
        \caption{The azimuthally-averaged toroidal magnetic field along the polar direction at $r \approx 50$\,km.
        }
        \label{fig:mag}
      \end{center}
\end{figure}

Once a part of the mean magnetic field in the polar region plunges into the BH, the BH spin can further amplify the toroidal magnetic field by the winding.
This enhances the magnetic pressure at the polar region, and the magnetic field expands vertically to the disk.
Along with this, the rest-mass density at this region decreases due to the accretion into the BH or expansion due to the aforementioned magnetic pressure.
Eventually, a large-scale helical magnetic field around the spin axis is developed, and the magnetization parameter $b^2/4\pi\rho$ exceeds the unity forming a magnetosphere.
Since this system undergoes the prompt collapse, the rest-mass density along the BH axis after the merger is lower than the non-prompt collapse case due to the lack of the fallback matter, and thus, it is easier to develop a magnetosphere compared to the non-prompt collapse case~\cite{kiuchi2023jul}.
In this magnetically dominated region, the outgoing Poynting flux is launched and drives an outflow. 
By contrast, the Poynting flux-driven outflow was not launched until $t-t_\mathrm{merger}\approx 1.1$\,s in the non-prompt collapse case~\cite{kiuchi2023jul}. The magnetosphere developed is supported and collimated by the gas pressure from the matter.

\begin{figure}[t]
      \begin{center}
        \includegraphics[scale=0.25]{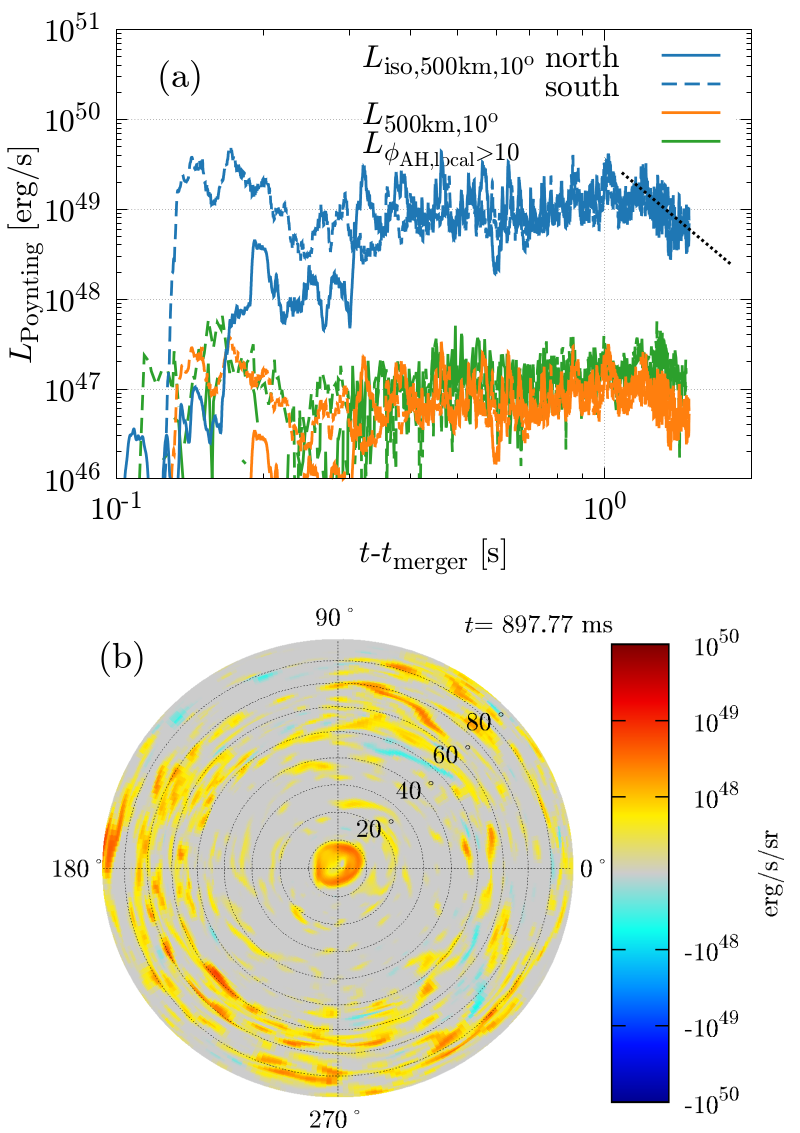} \\
        \caption{The quantities related to the magnetically-driven outflow. 
        The top panel plots the Poynting luminosity $L_\mathrm{500km,10^{\circ}}$ and isotropic-equivalent luminosity $L_\mathrm{iso,500km,10^{\circ}}$ evaluated at $r\approx500$\,km and opening angle with $10^{\circ}$.
        It also plots the Poynting luminosity $L_\mathrm{\phi_\mathrm{AH,local}>10}$ evaluated at the region with $\phi_\mathrm{AH,local}>10$.
        The dotted line emphasizes the decreasing trend of the Poynting luminosity for $t-t_\mathrm{merger}\gtrsim1\mathrm{\,s}$.
        The bottom panel plots the angular distribution of the Poynting luminosity per steradian evaluated on the sphere at $r\approx500$\,km for the north hemisphere.
        See also the animation~\cite{pfanim}.
        }
        \label{fig:signal}
      \end{center}
\end{figure}

Figure~\ref{fig:signal} (a) plots the isotropic-equivalent Poynting luminosity evaluated on the sphere at $r\approx500\mathrm{\,km}$ and opening angle of $10^{\circ}$ for the north and south hemispheres.
It shows that the luminosity reaches $L_\mathrm{iso,500km,10^{\circ}}\approx10^{49}\mathrm{\,erg/s}$ at $t-t_\mathrm{merger}\approx0.3\mathrm{\,s}$ and $0.13\mathrm{\,s}$ for the north and south hemispheres, respectively. 
Since the magnetosphere is formed after at least $\approx3$\,periods of the $\alpha\Omega$ dynamo cycle, it contributes to the launch of the Poynting flux.
We found that this Poynting luminosity is powered by the Blandford-Znajek mechanism~\cite{blandford1977} by magnetic fields penetrating a rapidly spinning BH.
Figure~\ref{fig:signal} (a) also plots the Poynting luminosity $L_\mathrm{500km,10^{\circ}}$ confined in a region within $\theta < 10^{\circ}$ at $r \approx 500~\rm{km}$ and the Poynting luminosity $L_\mathrm{\phi_\mathrm{AH,local}>10}$ evaluated at the region with $\phi_\mathrm{AH,local}>10$ on the apparent horizon. 
Here, $\phi_\mathrm{AH,local}$ is the local MADness parameter~\cite{hayashi2023jun} defined as
$\phi_\mathrm{AH,local}:=\left| \mathcal{B}^{r} / \sqrt{ \rho_{*} v^{r} c }  \right|$.
They exhibit essentially the same time evolution, which indicates that the magnetically dominated region on the BH provides the Poynting flux in the magnetosphere.

Figure~\ref{fig:signal} (b) plots the angular distribution of the Poynting luminosity per steradian evaluated on the sphere at $r\approx500$\,km.
It shows that the ring-shaped Poynting flux with $\approx10^{49}\mathrm{\,erg/s/str}$ exists along the BH spin axis with a collimation angle of $\alt10^{\circ}$.
The launching of the Poynting flux is in great contrast to the previously reported prompt collapse case~\cite{Ruiz2017oct}. 
We deduce that the discrepancy may be attributed to the shortness of their simulation of $t-t_\mathrm{merger}\lesssim 0.1$\,s and/or the different BNS model.

Figure~\ref{fig:signal} (a) shows that for $t-t_\mathrm{merger}\agt1\mathrm{\,s}$ the isotropic-equivalent Poynting luminosity starts decreasing from $10^{49}\mathrm{\,erg/s}$.
This is caused likely by the post-merger mass ejection~\cite{hayashi2022jul,hayashi2023jun}: 
The disk expands as the mass ejection proceeds, and the gas pressure around the funnel region decreases.
The magnetic pressure in the magnetosphere remains because the magnetic flux does not significantly dissipate in the ideal magnetohydrodynamics.
Since the collimated funnel structure of the magnetosphere is maintained by the balance of the magnetic pressure in the magnetosphere and the gas pressure from the disk, the opening angle of the magnetosphere increases. 
As the opening angle increases~\cite{Pais2023mar,Pais2024jul_arxiv}, the collimation of the magnetic flux becomes loosened, and the Poynting flux gradually decreases (see the animations~\cite{2Danim,pfanim}).
We expect that in a few $\mathrm{s}$, a substantial fraction of the disk matter will be ejected, and then, the magnetically driven Poynting flux will damp significantly.

{\it Gravitational wave, neutrino, and kilonova}.--
In addition to the Poynting flux-driven outflow, we also analyze the GW emission, neutrino emission, and ejection of neutron-rich matter in this simulation.

\begin{figure*}[t]
      \begin{center}
        \includegraphics[scale=0.2]{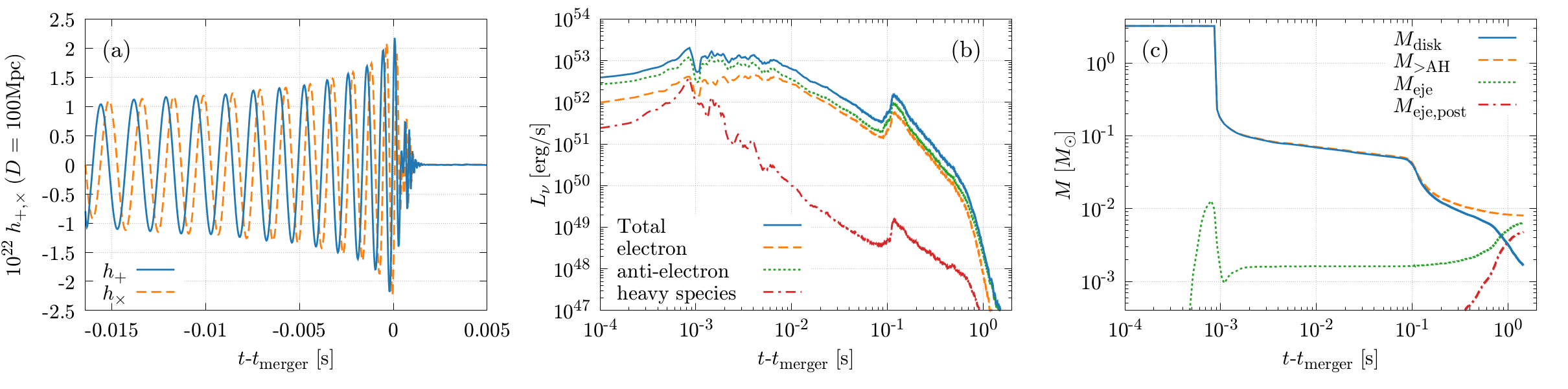} \\
        \caption{The quantities related to multimessenger signals.
        Left: gravitational waveform for $l=m=2$ mode.
        Middle: neutrino luminosity for the electron, anti-election, heavy lepton, and the total. 
        Right: the baryonic mass of the matter remaining outside the apparent horizon (dashed), the disk (solid), the total ejecta (dotted), and the post-merger ejecta (dashed-dotted).
        }
        \label{fig:signal_gnk}
      \end{center}
\end{figure*}

Figure~\ref{fig:signal_gnk} (a) plots the gravitational waveform for the $l=m=2$ mode characterized by the inspiral and subsequent quasinormal ringdown~\cite{hotokezaka2013}.
Figure~\ref{fig:signal_gnk} (b) plots the neutrino luminosity for the electron, anti-electron, heavy-lepton, and total.
The luminosity becomes maximum after the merger with $\approx10^{53}\mathrm{\,erg/s}$.
At $t-t_\mathrm{merger}\approx0.1\mathrm{\,s}$, the luminosity shows another peak since the MRI started to be fully resolved in the disk and the viscous heating driven by the MRI turbulent enhances the temperature (see Supplement Material).
After this, the disk expansion and accretion onto the BH proceed, and the luminosity decreases due to the temperature decrease because the neutrino energy emission rate is approximately proportional to $T^6$~\cite{fuller1985jun}.
As the neutrino luminosity steeply drops, the energy generated by the turbulent viscous heating can then be used for disk expansion and induces post-merger mass ejection~\cite{fernamdez2013aug,just2015feb,fujibayashi2020apr,fujibayashi2020dec,just2022jan,Fujibayashi2023jan}. 
Figure~\ref{fig:signal_gnk} (c) plots the baryonic mass of the remnant, disk, total ejecta, and post-merger ejecta.
During the merger stage, the dynamical ejecta with its mass $\approx1.6 \times 10^{-3} {\,M_{\odot}}$ is launched.
At $t-t_\mathrm{merger}\approx0.1\mathrm{\,s}$ disk mass starts decreasing steeply due to the MRI-driven turbulent viscosity, and post-merger mass ejection sets in at $t-t_\mathrm{merger}\approx0.4\mathrm{\,s}$. 
At the end of the simulation, the disk mass decreases to $\approx1.6\times10^{-3}\,M_{\odot}$ and the post-merger ejecta reaches $4.7\times10^{-3}\,M_{\odot}$, which is $\approx8\%$ of the disk mass at $t-t_\mathrm{merger}\approx0.02\mathrm{\,s}$, making a total ejecta mass $6.3\times10^{-3}\,M_{\odot}$. 
This indicates that the kilonova light curve would be similar to that of the BNS merger leaving a short-lived hypermassive neutron star (HMNS) presented in Refs.~\cite{Fujibayashi2023jan,Kawaguchi:2023zln} and is too faint to explain the observation of AT2017gfo.

{\it Summary and discussion}.--
We performed the first numerical-relativity simulation for prompt-collapse BNS merger which included neutrino radiation transfer and magnetohydrodynamics effects.
The simulation lasts for $\approx1.5\mathrm{\,s}$ after the merger, the longest among BNS merger simulations that self-consistently solve from the inspiral to the late post-merger stage. 
We focused on an asymmetric BNS merger so that the dynamical mass ejection and the massive disk formation proceed during the merger stage.
We found that the gravitational waveforms, neutrino luminosity, and mass ejection are consistent with the previous simulations~(e.g., Refs.~\cite{hotokezaka2013,kiuchi2023jul}).

We found that prompt-collapse BNS mergers can develop the magnetosphere which launches a Poynting flux-driven outflow like the BNS merger that has a long-living HMNS as a remnant~\cite{kiuchi2024mar} or BHNS merger~\cite{hayashi2022jul,hayashi2023jun}.
The MRI-driven $\alpha\Omega$ dynamo activity in the disk amplifies the magnetic field and provides an aligned magnetic field toward the polar region.
Then, the BH spin further amplifies the magnetic field, generating a large-scale magnetic field due to the tower effect.
It requires more than $0.1\mathrm{\,s}$ after the merger until the magnetosphere is developed and the isotropic-equivalent Poynting flux reaches as high as $10^{49}\mathrm{\,erg/s}$. 
At $t-t_\mathrm{merger}\approx 1\,\mathrm{s}$ the isotropic-equivalent Poynting flux starts decreasing due to the increase of the opening angle of the magnetosphere, which is induced by the post-merger mass ejection (see Supplement Material).

We have to note that in the present simulation, the maximum value of the initial magnetic field strength is set to be $10^{15}\mathrm{\,G}$, which is as high as that for magnetars, i.e., much higher than observed in binary pulsars~\cite{lorimer2008nov}.  This treatment is motivated by an expectation that if the MRI is resolved well, the overall post-merger evolution processes remain essentially unchanged irrespective of the initial magnetic field strength. 
However, in the simulation with an initially high magnetic field, not only the MRI but also the magnetic winding may play a role in enhancing the field strength in the remnant disk. 
The butterfly diagram of the averaged toroidal magnetic field shows that the aligned magnetic field at the polar region is generated not only by the periodic magnetic field amplification due to the $\alpha\Omega$ dynamo but also by the winding of the magnetic field.
We expect that in the system of a realistic magnetic field strength, MRI-driven $\alpha\Omega$ dynamo activity will solely provide the magnetic field for magnetosphere formation. 
The higher-resolution simulation starting from a weaker pre-merger magnetic field should be conducted as future work.

{\it Acknowledgments}.--
We thank the members of the Computational Relativistic Astrophysics group in AEI for useful discussions.
We also thank Alexis Reboul-Salze for the helpful discussion.
The numerical simulation of this work was performed on the computational resources of the supercomputer Fugaku provided by RIKEN through the HPCI System Research Project (Project ID: hp220174, hp220392, hp230084, and hp240039). 
This work was supported by Grant-in-Aid for Scientific Research (20H00158, 23H04900, 23H01172, and 23K25869) of JSPS/MEXT.

\bibliography{paper}

\clearpage
\newpage
\onecolumngrid
\begin{center}
{\large \bf Supplement Material}
\vspace*{1cm}
\twocolumngrid
\end{center}

\section{Dynamical mass ejection}
\begin{figure*}[p]
      \begin{center}
        \includegraphics[scale=0.11]{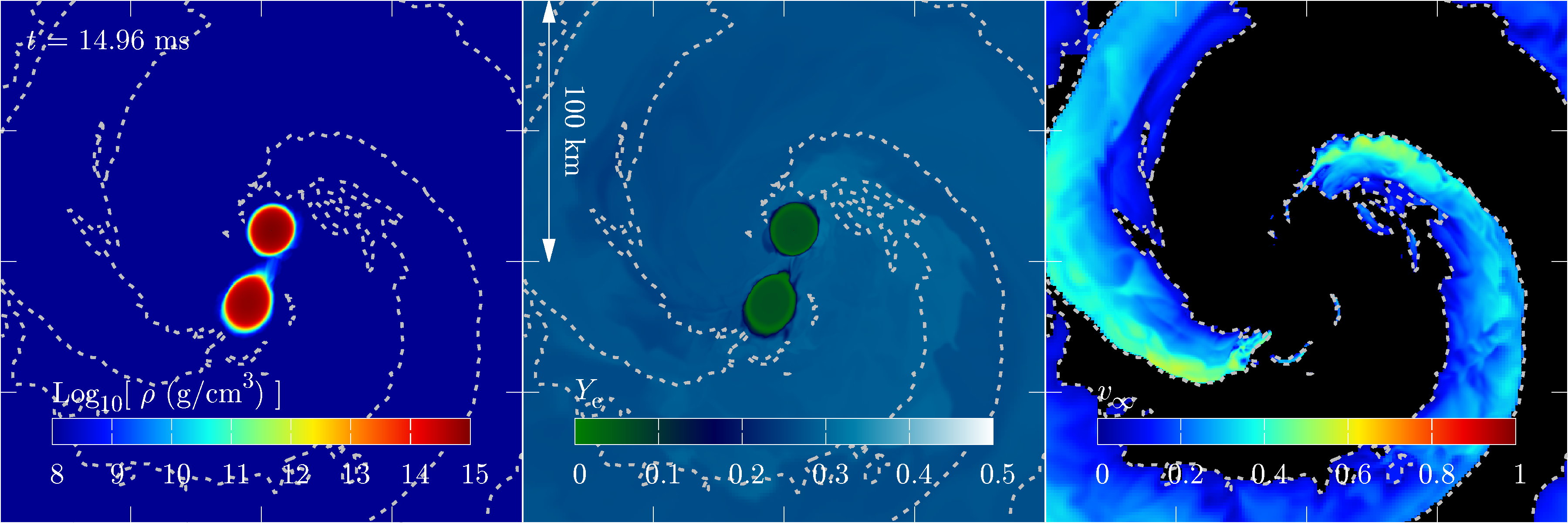} \\
        \vspace{2mm}
        \includegraphics[scale=0.11]{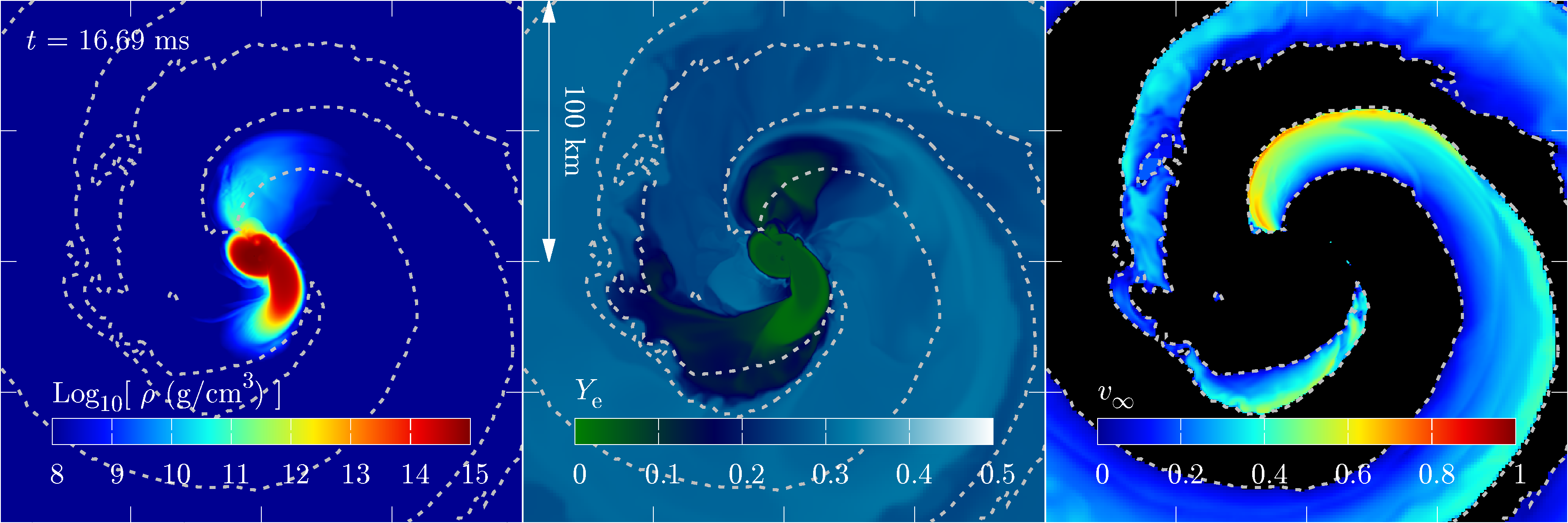} \\
        \vspace{2mm}
        \includegraphics[scale=0.11]{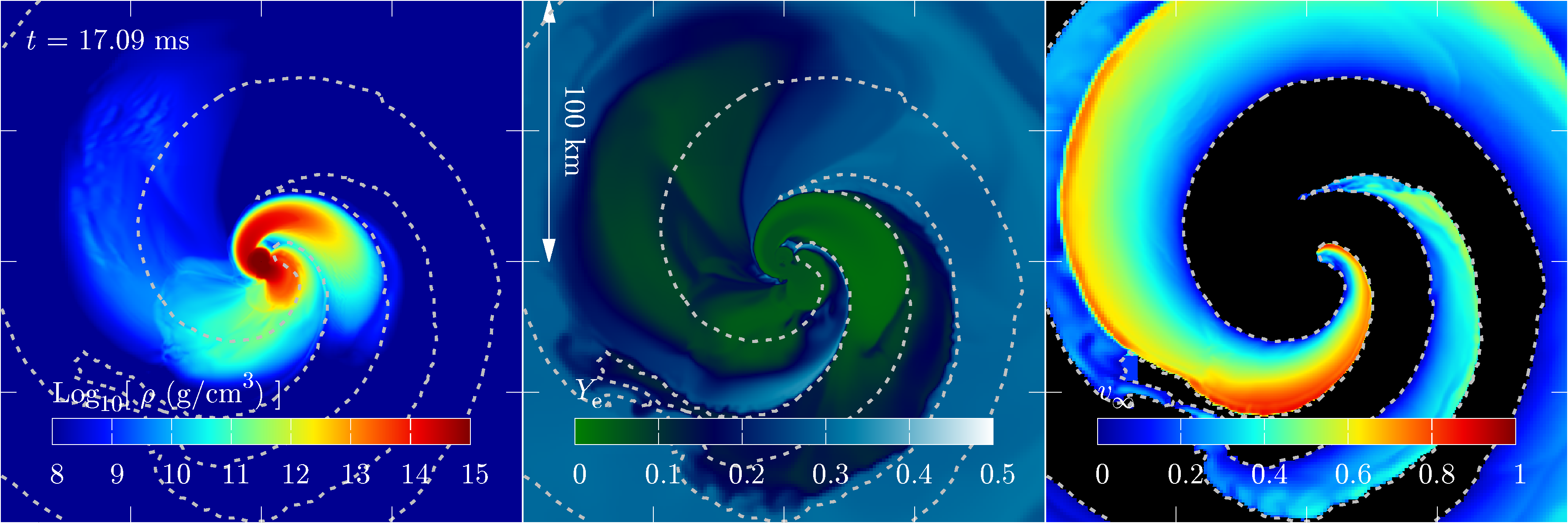} \\
        \vspace{2mm}    
        \includegraphics[scale=0.11]{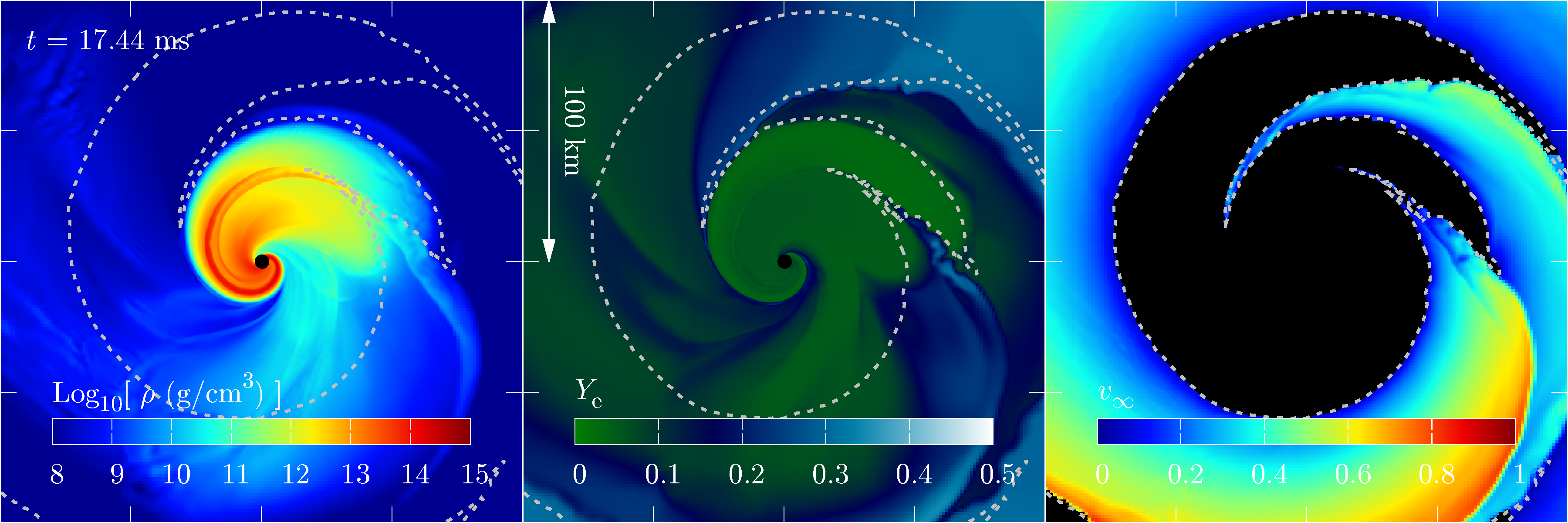} \\
        \caption{The 2D snapshots on the $x$-$y$ plane with a region of  $[-100~{\rm km}:100~{\rm km}]$ for both $x$ and $y$ directions at $t-t_\mathrm{merger}\approx-1.4,\,0.3,\,0.7$, and 1.4\,ms from the top to the bottom panel, respectively. The panels show the rest-mass density $\rho$\,$\mathrm{(g/cm^3)}$ (left), the electron fraction $Y_\mathrm{e}$ (center), and the terminal velocity of the unbound component $v_{\infty}$ (right), respectively. The black-filled circle indicates the interior of the apparent horizon, and the unbound component is enclosed by the gray dashed curve. 
        }
        \label{fig_sup:xysnapshot}
      \end{center}
\end{figure*}

To show the dynamical mass ejection process, 
in Fig.~\ref{fig_sup:xysnapshot}, we plot the 2D snapshots on the $x$-$y$ plane for the rest-mass density $\rho$\,$\mathrm{(g/cm^3)}$, the electron fraction $Y_\mathrm{e}$, and the terminal velocity $v_{\infty}$.
The terminal velocity is defined by
\begin{eqnarray}
v_{\infty} &:=& \sqrt{1-\Gamma_{\infty}^{-2}}, \nonumber \\
\Gamma_{\infty} &:=& -\frac{hu_{t}}{h_{\mathrm{min}}(Y_\mathrm{e})},
\end{eqnarray}
where $h$ is the specific enthalpy and $h_\mathrm{min}(Y_\mathrm{e})$ denotes the minimum value of it for a given electron fraction $Y_\mathrm{e}$. 
The ejecta are composed of two different components.
One component is driven by the tidal force of the heavier neutron star, by which the lighter neutron star is elongated, and a fraction of the elongated matter gains sufficient kinetic energy and then becomes unbound~\cite{hotokezaka2013jan}. 
Mass ejection by this process is typically seen in the black hole-neutron star merger with tidal disruption. 
Ejecting this component does not involve a thermal process, and thus, the chemical composition is not significantly modified from the original neutron star~\cite{Wanajo:2022jgw}. 
However, this is not a major component because the mass ratio in this case, $Q=M_1/M_2\approx 1.32$, is smaller than that in the typical black hole-neutron star merger case $Q=M_\mathrm{BH}/M_\mathrm{NS}$, and thus the tidally driven dynamical ejecta becomes less massive~\cite{foucart2019may,hayashi2021feb}.
The other component originates from the contact region of two neutron stars.
This component is likely to be driven by the combination of the shock heating and the swing-by effect caused by the orbital motion of the heavier neutron star.
Since the thermal process plays an important role, the chemical composition of this component is modified from the original neutron star to less neutron-rich one. 
In the present simulation, this second component is the major part of the dynamical ejecta, as can be seen in the $Y_\mathrm{e}$ histogram of the ejecta showing that $Y_\mathrm{e}$ widely ranges from $\approx 0.03$ to $\approx 0.3$ (see also Fig.~\ref{fig_sup:ejecta}(a)).
This component had mass of $\sim10^{-2}\,M_{\odot}$ right after the merger, but it rapidly damps to $\sim 2 \times 10^{-3}M_\odot$ because soon after becoming unbound, its propagation is disturbed by the deformed lighter neutron star (see also Fig. 4(c) in the main text at $t-t_\mathrm{merger} \approx 10^{-3}~{\rm s}$.).

\section{Magnetic field amplification}

\begin{figure*}[t]
      \begin{center}
        \includegraphics[scale=0.2]{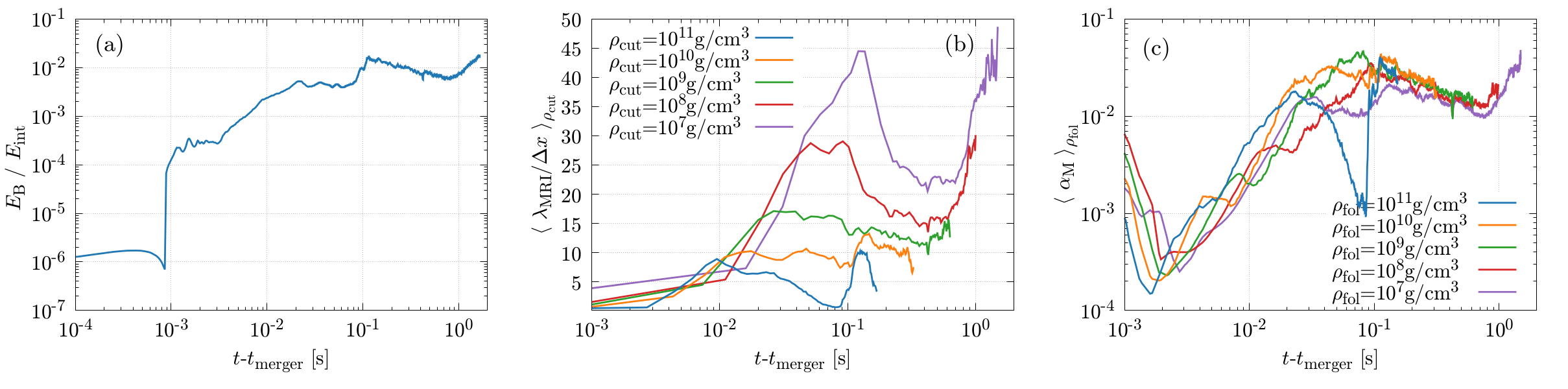} \\
        \caption{The time evolution of the quantities related to the magnetic field amplification.
        The left panel shows the ratio of the electromagnetic energy to the internal energy.
        The middle panel shows the volume-averaged MRI quality factor for selected cutoff rest-mass densities. The right panel shows the volume-averaged Shakura-Sunyaev $\alpha_\mathrm{M}$ parameter for the selected foliation of the rest-mass density.
        }
        \label{fig_sup:mag}
      \end{center}
\end{figure*}

Figure~\ref{fig_sup:mag}~(a) plots the ratio of the electromagnetic field energy to the internal energy $E_\mathrm{B}/E_\mathrm{int}$.
The sharp increase of $E_\mathrm{B}/E_\mathrm{int}$ at $t-t_\mathrm{merger}\approx10^{-3}$\,s is mainly due to the formation of the black hole, which reduces the total internal energy outside the apparent horizon.
$E_\mathrm{B}/E_\mathrm{int}$ kept on increasing for $t-t_\mathrm{merger}\alt10^{-1}$\,s due to the magnetic field winding associated with the differential rotation of the disk~\cite{shibata2006}, and non-axisymmetric MRI~\cite{balbus1998}.
Unlike the case with a hypermassive neutron star remaining after the merger~\cite{Kiuchi2014aug,kiuchi2023jul,kiuchi2024mar,Aguilera2020nov,Aguilera2022feb,Palenzuela2022jul,Aguilera2023nov}, the Kelvin-Helmholtz instability plays only a minor role in the magnetic field amplification since the shear layer between the two neutron stars developed at the onset of the merger is instantaneously swallowed into the black hole.
Figure~\ref{fig_sup:mag}~(b) plots the volume averaged MRI quality factor 
\begin{eqnarray}
\langle \lambda_{\mathrm{MRI,}z}/\Delta x \rangle_{\rho_\mathrm{cut}} := \frac{\displaystyle \int_{\rho>\rho_\mathrm{cut}}{\lambda_{\mathrm{MRI,}z}/\Delta x \, dx^3}}
{\displaystyle \int_{\rho>\rho_\mathrm{cut}}{dx^3}}.
\end{eqnarray}
Here, the wavelength of the fastest growing mode of the axisymmetric MRI is defined as $\lambda_{\mathrm{MRI,}z}:=2 \pi b_z / \sqrt{4 \pi \rho h + b^2} \Omega$.
It shows that the fastest growing mode of the axisymmetric MRI~\cite{balbus1991,balbus1998} starts being resolved at $t-t_\mathrm{merger}\approx10^{-2}$\,s for the region with $\rho\alt10^{10}\mathrm{\,g/cm^3}$ and the MRI in the high-density part of the disk with $\rho\agt10^{11}\mathrm{\,g/cm^3}$ starts being resolved for $t-t_\mathrm{merger}\gtrsim 10^{-1}$\,s.
As the MRI becomes resolved fully in the disk, the electromagnetic energy experiences an additional increase, and it saturates at $\approx1\%$ of the internal energy at $t-t_\mathrm{merger}\approx 10^{-1}$\,s~\cite{hayashi2022jul,hayashi2023jun,kiuchi2023jul}.
Figure~\ref{fig_sup:mag}~(c) plots the volume averaged Shakura-Sunyaev $\alpha_\mathrm{M}$ parameter 
\begin{eqnarray}
\langle \alpha_\mathrm{M} \rangle_{\rho_\mathrm{fol}} := \frac{\displaystyle \int_{10\rho_\mathrm{fol}>\rho>\rho_\mathrm{fol}}{\alpha_\mathrm{M} \, dx^3}}
{\displaystyle \int_{10\rho_\mathrm{fol}>\rho>\rho_\mathrm{fol}}{dx^3}},
\end{eqnarray}
where $\alpha_\mathrm{M}:= -b^r b_\varphi / 4 \pi P$.
It shows that once the MRI is activated in the disk, $\alpha_\mathrm{M}$ takes a value of $1$--$3\times 10^{-2}$, representing that the turbulent state is established and the effective viscosity plays an essetial role in the disk evolution. 

\section{2D snapshots}

\begin{figure*}[p]
      \begin{center}
        \includegraphics[scale=0.08]{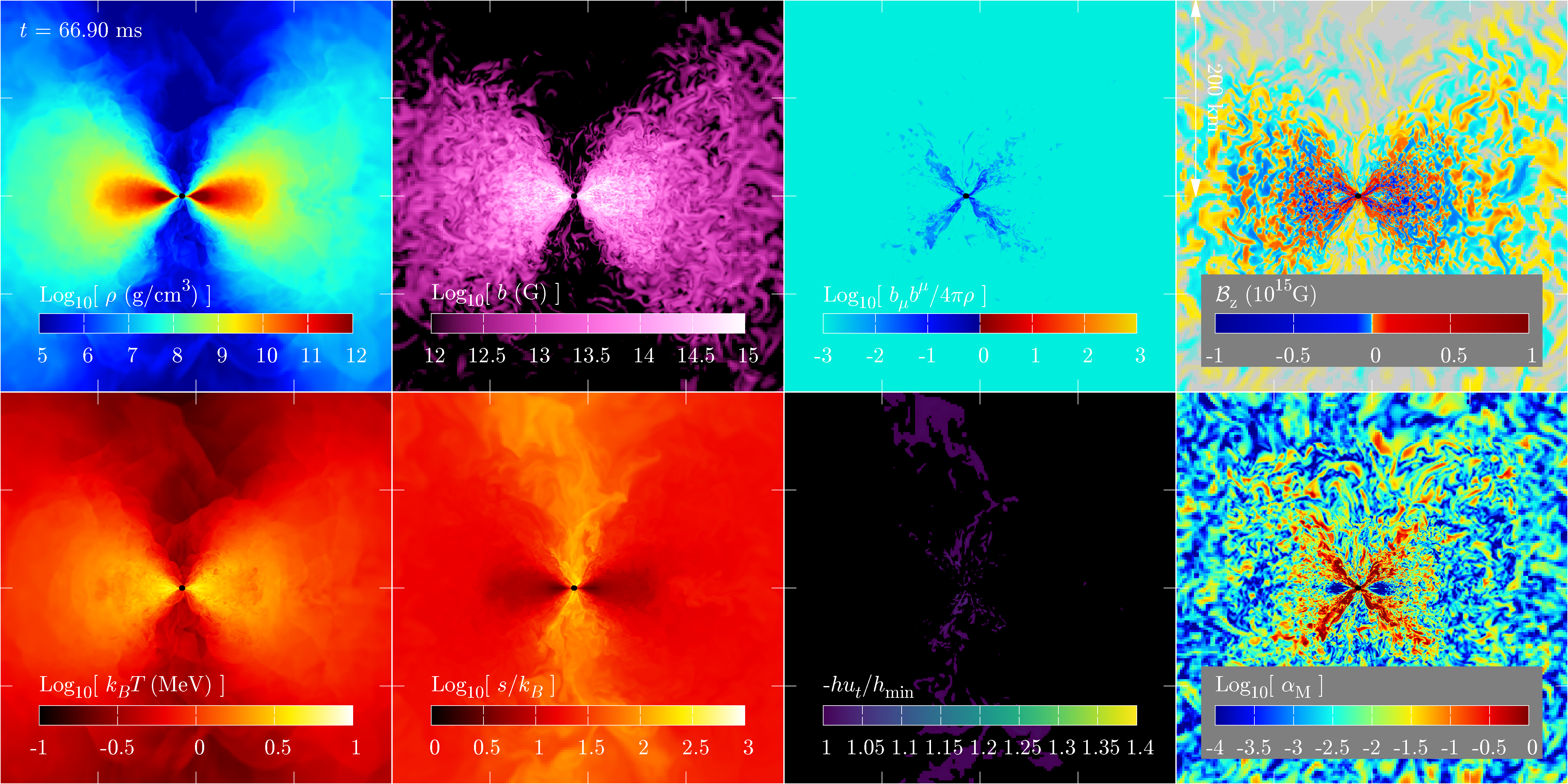} \\
        \vspace{2mm}
        \includegraphics[scale=0.08]{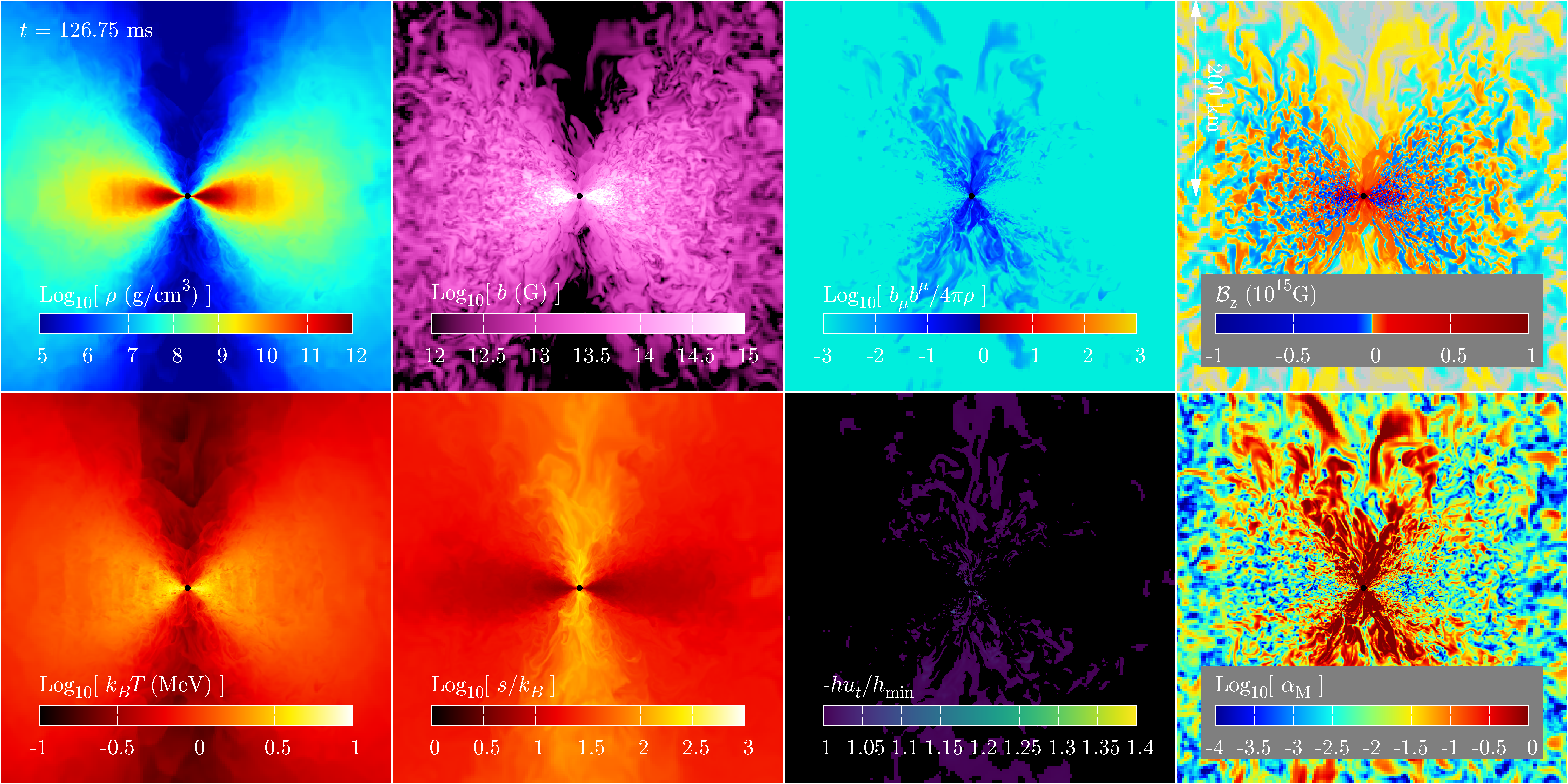} \\
        \vspace{2mm}
        \includegraphics[scale=0.08]{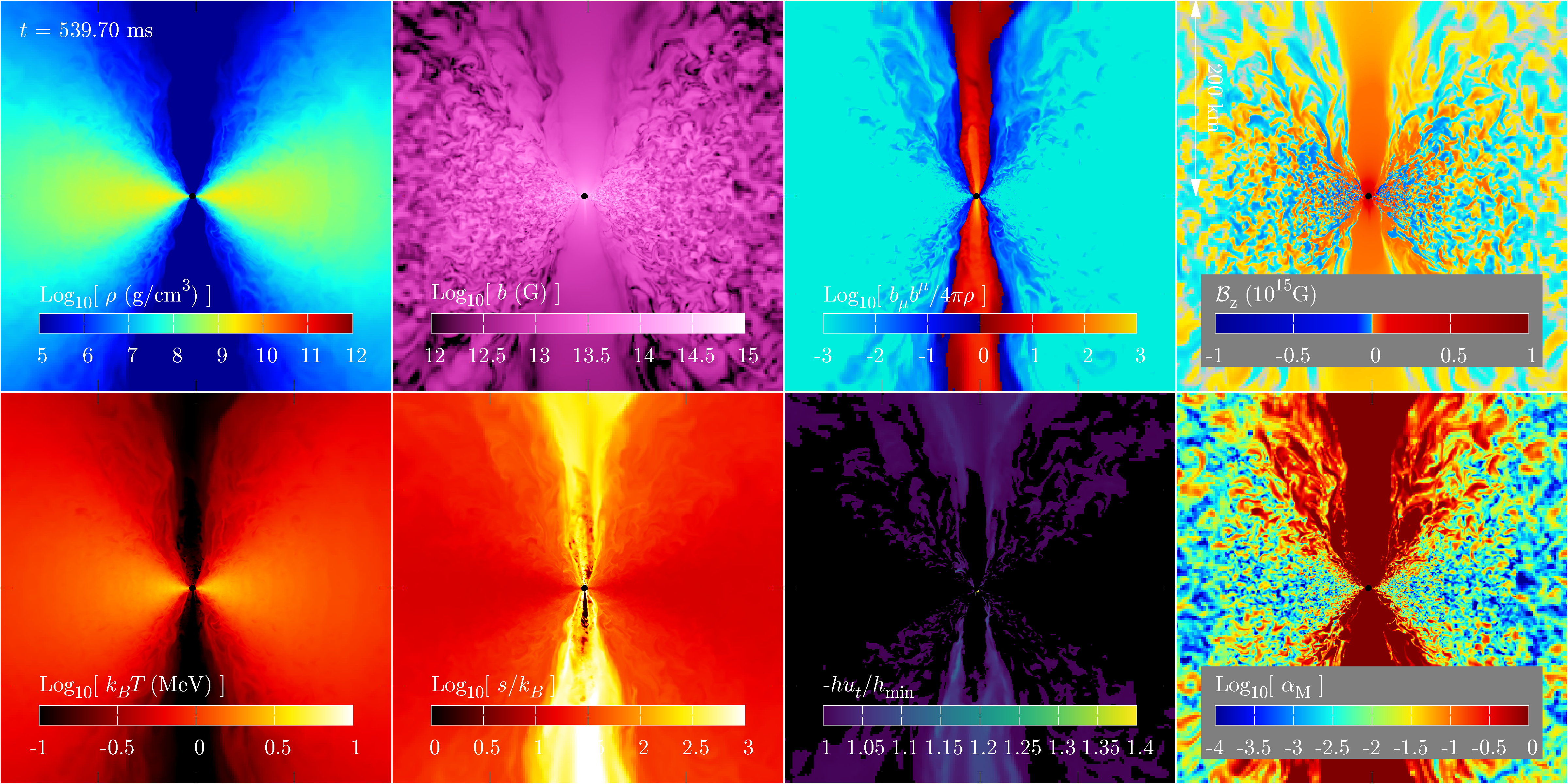} \\
        \caption{The 2D snapshots on the $x$-$z$ plane with a region of  $[-200~{\rm km}:200~{\rm km}]$ for both $x$ and $z$ directions at $t-t_\mathrm{merger}\approx0.05,\,0.11$, and $0.52$\,s from the top to the bottom panels, respectively. The panels show the rest-mass density $\rho$\,$\mathrm{(g/cm^3)}$, the magnetic field strength $b$, the magnetization parameter $b^2/4\pi\rho$, the $z$-component of the magnetic field in the laboratory frame $\mathcal{B}_z$, the temperature $T\,\mathrm{(MeV)}$, the entropy per baryon $s$ in units of $k_B$, the criterion for unbound matter $-hu_{t}/h_{\mathrm{min}}(Y_\mathrm{e})$, and the Shakura-Sunyaev $\alpha_\mathrm{M}$ parameter from the top-left to the bottom-right panel, respectively. See also for animation~\cite{2Danim}.
        }
        \label{fig_sup:xzsnapshot}
      \end{center}
\end{figure*}

Figure~\ref{fig_sup:xzsnapshot} displays the 2D snapshots on the $x$-$z$ plane for the rest-mass density $\rho$\,$\mathrm{(g/cm^3)}$, the magnetic field strength $b$, the magnetization parameter $b^2/4\pi\rho$, the $z$-component of the magnetic field in the laboratory frame $\mathcal{B}_z$, the temperature $T\,\mathrm{(MeV)}$, the entropy per baryon $s$ in units of $k_B$, the criterion for unbound matter $-hu_{t}/h_{\mathrm{min}}(Y_\mathrm{e})$, and the Shakura-Sunyaev $\alpha_\mathrm{M}$ parameter associated with the $r$-$\varphi$ component of the Maxwell stress for the time slice at $t-t_\mathrm{merger}\approx0.05,\,0.11$, and $0.52$\,s.

\section{Ejecta properties}
\begin{figure*}[t]
      \begin{center}
        \includegraphics[scale=0.25]{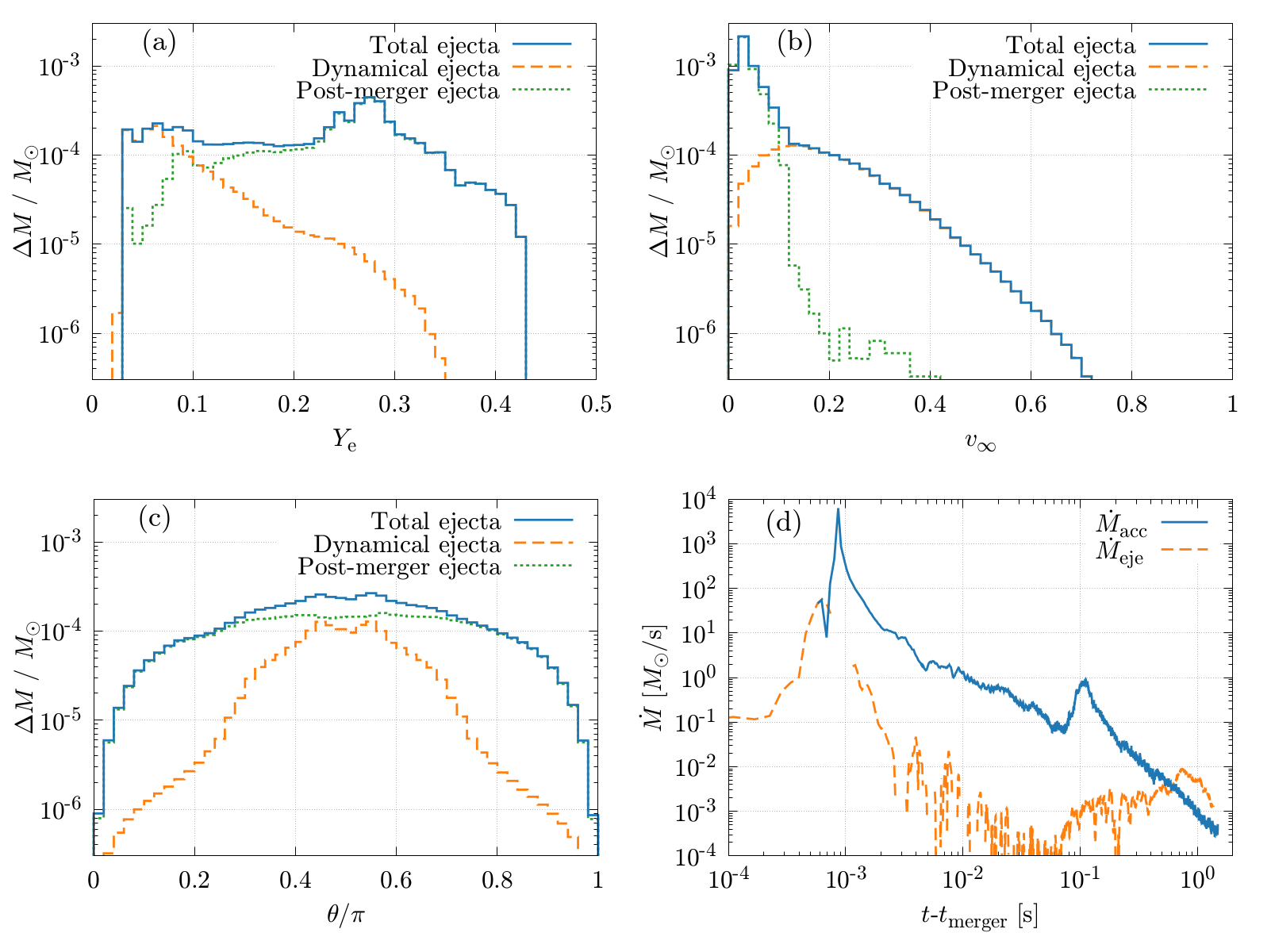} \\
        \caption{The panels (a)--(c) plot the mass histogram of the ejecta as a function of electron fraction, terminal velocity, and polar angle, respectively.
        The histogram for the total (solid), dynamical (dashed), and post-merger ejecta (dotted) are shown.
        The panel (d) shows the time evolution of mass accretion rate onto the black hole (solid) and mass ejection rate (dashed).}
        \label{fig_sup:ejecta}
      \end{center}
\end{figure*}

Figure~\ref{fig_sup:ejecta} (a)--(c) plot the mass histogram of the ejecta as a function of electron fraction, terminal velocity, and polar angle, respectively.
The dynamical ejecta is evaluated at $t-t_\mathrm{merger}\approx0.04$\,s, and the total ejecta is evaluated at the end of the simulation.
The post-merger ejecta is obtained by subtracting the dynamical part from the total.
The $Y_e$ distribution of dynamical ejecta peaks at $\approx0.05$ and extends to $\approx0.3$.
This is because the dynamical ejecta consists of tidally driven components, which do not experience thermal process and preserve the original neutron star property, and the shock-heating accelerated component.
The post-merger ejecta peaks at $\approx0.27$, and this profile is similar to our previous short-lived remnant hypermassive neutron star case~\cite{kiuchi2023jul}. 
The velocity distribution of the dynamical ejecta peaks at $0.13c$ and extends to the mildly relativistic regime.
The post-merger ejecta peaks at $\approx0.03$c. It is worth pointing out that the high-velocity component is present up to $\sim 0.7c$ in contrast to black hole-neutron star mergers for which the maximum velocity is always $\sim 0.4c$~\cite{Chen:2024ogz}. 
The polar angle distribution of the dynamical ejecta is concentrated around the equatorial plane and extends from $\approx0.3\pi$ to $\approx0.7\pi$.
The post-merger ejecta extends more from $\approx0.1\pi$ to $\approx0.9\pi$. We note that the concentration of the dynamical ejecta around the equatorial plane would be modified if the heating due to the radioactive decay of $r$-process elements are taken into account~\cite{Kawaguchi:2024hdk}. 

Figure~\ref{fig_sup:ejecta} (d) plots the rates of the mass accretion onto the black hole and mass ejection.
The mass accretion rate is defined by
\begin{eqnarray}
  \dot{M}_{\mathrm{AH}}&:=&\oint_{\mathrm{AH}} \rho \sqrt{-g} u^{i} d S_{i} \nonumber \\
  &\approx&\oint_{r=\max{(r_{\mathrm{AH}}})} \rho_{*} v^r r^2\sin{\theta} d\theta d\varphi, 
\end{eqnarray}
and the mass ejection rate is defined by the time derivative of the ejecta mass defined by 
\begin{eqnarray}
  M_{\rm eje}&:=&\int_{-hu_{t}>h_{\rm min},r>r_{\rm AH}}\rho_*d^3x + M_{\rm esc}, \\
  \dot{M}_{\rm esc}&:=&\oint \rho \sqrt{-g} u^{i} dS_{i}, \\
  M_{\rm esc}&:=&\int^{t} \dot{M}_{\rm esc} dt.
\end{eqnarray}
The surface integral for $\dot{M}_{\rm esc}$ is performed at the outer boundary of the $l=3$ refinement level.
The peak of the mass accretion rate at $t-t_\mathrm{merger}\lesssim10^{-3}$\,s corresponds to the formation of the black hole.
Once the fastest growing mode of the MRI starts being resolved, the accretion rate shows an additional peak at $t-t_\mathrm{merger}\approx0.1$\,s due to the accretion induced by the turbulent effective viscosity.
If the MRI were resolved from the beginning of the simulation, we expect that there might be no peak here but have a slightly higher mass accretion rate for $t-t_\mathrm{merger}\lesssim0.1$\,s.
The mass ejection rate peaks at $t-t_\mathrm{merger}\approx10^{-3}$\,s which corresponds to the dynamical mass ejection.
Then it starts increasing again at $t-t_\mathrm{merger}\gtrsim0.1$\,s indicating the onset of the post-merger mass ejection facilitated by the MRI-driven turbulent viscosity, and it exceeds the mass accretion rate at $t-t_\mathrm{merger}\gtrsim0.6$\,s.
Approximately at the same time as this excess, neutrino luminosity starts dropping steeply (see also Fig.~4(b) in the main text.).
This corresponds to the fact that neutrino cooling became inefficient, and the thermal energy generated by the turbulent effective viscous heating can then be efficiently used for the disk expansion and post-merger mass ejection.

\section{Evolution of the magnetically dominated region}
\begin{figure*}[t]
      \begin{center}
        \includegraphics[scale=0.2]{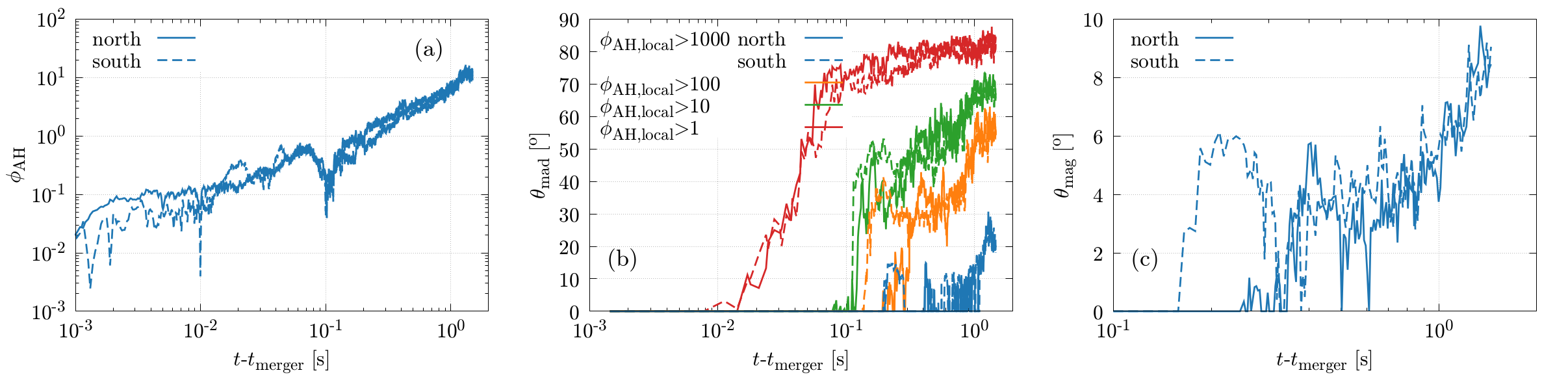} \\
        \caption{The time evolution of the quantities related to the magnetically dominated region.
        The left panel plots the global MADness parameter $\phi_\mathrm{AH}$.
        The middle panel plots the opening angle of the region with the local MADness parameter $\phi_\mathrm{AH,local}>1,\,10,\,100,\,$ and 1000.
        The right panel plots the opening angle of the magnetosphere.
        }
        \label{fig_sup:ahmag}
      \end{center}
\end{figure*}

Figure~\ref{fig_sup:ahmag}~(a) plots the MADness parameter defined by~\cite{tchekhovskoy2011}
\begin{eqnarray}
  \phi_{\mathrm{AH}} := \frac{ \Phi_{\mathrm{AH}} }{ \sqrt{ \dot{M}_{\mathrm{AH}} r_{\mathrm{AH}}^2 c } },
\end{eqnarray}
where
\begin{eqnarray}
  \Phi_{\mathrm{AH}}&:=&\oint_{\mathrm{AH}} B^i \sqrt{\gamma} d S_{i} \nonumber \\
  &\approx&\oint_{r=\max{(r_{\mathrm{AH}}})} \mathcal{B}^r r^2 \sin{\theta} d\theta d\varphi,  
\end{eqnarray}
for the north and south hemispheres. 
As the mass accretion rate decreases, the MADness parameter increases, but it does not reach $50$, which is considered to be the criterion for the magnetically arrested disk (MAD)~\cite{tchekhovskoy2011}.
Though our system does not reach the global MAD state, the Poynting flux-driven outflow is launched.
We consider this to be due to the highly magnetized region developing locally near the black hole spin axis while mass accretion dominance is sustained near the equatorial plane. 

Figure~\ref{fig_sup:ahmag} (b) shows the opening angle of the region with the local MADness parameter $\phi_\mathrm{AH,local}>1,\,10,\,100,\,$ and 1000 for the north and south hemisphere. 
The launching of the Poynting flux-driven outflow corresponds to the emergence of the region with the $\phi_\mathrm{AH,local}>100$, which is consistent with the black hole-neutron star merger case~\cite{hayashi2023jun}.

Figure~\ref{fig_sup:ahmag} (c) plots the opening angle of the magnetosphere defined by 
\begin{eqnarray}
\theta_\mathrm{mag}&:=&\arccos \left( 1-\frac{A_\mathrm{mag}}{2\pi} \right) \frac{180}{\pi},\nonumber \\
A_\mathrm{mag}&:=&\oint_{b^2/4\pi\rho>1}\sin{\theta} d\theta d\varphi,
\end{eqnarray}
at $r\approx500\mathrm{\,km}$ for the north and south hemispheres.
For $t-t_\mathrm{merger}\alt1\mathrm{\,s}$ the opening angle is $\theta_\mathrm{mag}\alt6^{\circ}$ and at $t-t_\mathrm{merger}\approx1.4\mathrm{\,s}$ the opening angle increases to $\theta_\mathrm{mag}\approx9^{\circ}$.
This results from the decrease in the gas pressure of the funnel region due to the post-merger mass ejection.
It indicates the Poynting-flux driven outflow would be preserved for $O(1)$\,s.

\end{document}